\documentstyle[epsf,12pt]{article}
\newcommand{\newsection}{    
\setcounter{equation}{0}
\section}
\def\appendix#1{
\addtocounter{section}{1}
\setcounter{equation}{0}
\renewcommand{\thesection}{\Alph{section}}
\section*{Appendix \thesection\protect\indent #1}
\addcontentsline{toc}{section}{Appendix \thesection\ \ \ #1}
}
\newcommand{\rf}[1]{(\ref{#1})}

\def\be{\begin{equation}}
\def\ee{\end{equation}}
\newcommand{\beq}{\begin{equation}}
\newcommand{\eeq}{\end{equation}}
\newcommand{\bea}{\begin{eqnarray}}
\newcommand{\eea}{\end{eqnarray}}

\newcommand{\tens}{{\!\otimes\!}}

\renewcommand{\l}{\lambda}

\newcommand{\oh}{\frac{1}{2}}

\newcommand{\Tr}{{\,\rm Tr}\:}

\newcommand{\eps}{\varepsilon}
\newcommand{\RA}{\right\rangle}
\newcommand{\LA}{\left\langle}
\newcommand{\e}{{\,\rm e}\,}

\newcommand{\hs}{\hspace{0.7cm}}

\def\e{{\,\rm e}\,}

\newcommand{\D}{{{\hbox{d}}}}
\begin{document}
\topmargin 0pt
\oddsidemargin 5mm
\headheight 0pt
\headsep 0pt
\topskip 9mm
\pagestyle{empty}
\hfill DTP-97-55 

\hfill SPhT-97/122

\hfill NBI-HE-97-51

\addtolength{\baselineskip}{0.20\baselineskip}
\begin{center}
\vspace{26pt}
{\large \bf {An Iterative Solution of the Three-colour Problem \\
\hspace{2.0cm}on a Random Lattice }}
\newline
\vspace{26pt}

{\sl B.\ Eynard}\hspace*{0.05cm}\footnote{
E-mail: 
Bertrand.Eynard@durham.ac.uk,
\\Permanent address: 
Service de Physique Th\'{e}orique de Saclay, 
F-91191 Gif-sur-Yvette Cedex, France
}
\\
\vspace{6pt}
Department of Mathematical Sciences \\
University of Durham, Science Labs. South Road \\
Durham DH1 3LE, UK\\

\vspace{18pt}
{\sl C. Kristjansen}\hspace{0.025cm}\footnote{E-mail:kristjan@alf.nbi.dk}
\\ 
\vspace{6pt}
The Niels Bohr Institute \\
 Blegdamsvej 17,
DK-2100 Copenhagen \O, Denmark \\
\end{center}
\vspace{20pt} 
\begin{center}
Abstract
\end{center}
We study the generalisation of Baxter's three-colour problem to a random 
lattice. Rephrasing the problem as a matrix model problem we discuss the 
analyticity structure and critical behaviour of the resulting matrix
model. Based on a set of loop equations we develop an algorithm which enables
us to solve the three-colour problem recursively.
\vfill{\noindent PACS codes: 05.20.y, 04.60.Nc }

\newpage 
\pagestyle{plain}
\setcounter{page}{1}

\newsection{Introduction \label{intro}}

 Consider a hexagonal lattice with ${\cal
 N}$ vertices. In how many ways is it possible to colour the links of
 the lattice with three colours, $A$, $B$ and $C$ so that no two links
 which meet at the same vertex carry the same colour. This is the
 classical three-colour problem solved by Baxter in
 1969~\cite{Baxter}. The problem is equivalent to a loop gas problem,
 namely the problem of enumerating the number of ways one can cover
 the lattice with closed, self-avoiding and non-intersecting loops
 which come in two different flavours, demanding
 that each vertex of the lattice belongs to a loop. This follows by
 noting that sequences of, say, $B$- and $C$-coloured links 
 form loops which exactly have the properties described
 above~\cite{Baxter}. The loop gas model, on the other hand, is a
 special version of the $O(2)$ model~\cite{Nie82}. Using the
 equivalence with the loop gas model Baxter furthermore proved that
 his problem was equivalent to the problem of colouring the faces of
 the hexagonal lattice with four different colours so that adjacent
 faces have different colours~\cite{Baxter}. The dual lattice of a
 hexagonal lattice is a triangular lattice. Baxter's three-colour problem
 can hence also be formulated as the problem of counting the number of
 ways of colouring the links of the triangular lattice so that the three
 sides of any triangle have three different colours. In addition, it has
 recently been shown that Baxter's three-colour problem is equivalent
 to the problem of counting the different foldings of the regular
 triangular lattice~\cite{DG94}.

 In this letter we  study the generalisation of Baxter's
 three-colour problem to a random triangular lattice. This can be
 viewed as coupling a hitherto unexplored type of matter to
 two-dimensional quantum gravity. Whereas the regular triangular
 lattice has six triangles meeting at each vertex, on a random
 triangular lattice any number of triangles can meet at a given
 vertex.  However, the random lattice three-colour problem shares many
 features with its regular lattice version. For instance,the random
 lattice three-colour problem is equivalent to a loop gas model,
 namely a version of the $O(2)$ model on a random lattice~\cite{Kos89}
 where only loops of even length are allowed. Furthermore, as in the
 regular lattice case, the random lattice three-colour problem is
 equivalent to the problem of counting the possible four-colourings of
the faces of the dual lattice. However, the random lattice three-colour
problem can not be given an interpretation as a folding problem. This 
requires that only configurations where an {\it even} number of
triangles meet at a given vertex are allowed. Finally, let us mention
that the three-colour problem on a random lattice has recently attracted
attention as a means of describing vertex models on random graphs~\cite{JP97}.

\newsection{The Model \label{sec1} }
The possible
 three-colourings of the random triangular lattice are generated by
 the following matrix model
\beq\label{ZABC}
Z(g)=\int_{N\times N}\, \D {\cal A}\, \D {\cal B}\, 
\D {\cal C}\, \exp\left\{-N \Tr\left({1 \over 2}
\left[{\cal A}^2 + 
{\cal B}^2 + {\cal C}^2\right]
 -\sqrt{g}\left[ {\cal ABC} + {\cal BAC} \right] \right)\right\}
\eeq
where ${\cal A}$, ${\cal B}$ and ${\cal C}$ are hermitian 
$N\times N$ matrices. More
precisely we have the following expansion of the free energy, 
$F(g)=\frac{1}{N^2}\log\, Z(g)$
\beq
F(g)=\sum_{h=0}^{\infty}N^{-2h} F_h(g),
\hspace{1.2cm} F_h(g)=\sum_v{\cal N}_h(2v) \,g^v
\label{count}
\eeq
where ${\cal N}_h(2v)$ is the number of closed, connected,
three-coloured random triangulations of genus $h$ consisting of $2v$
triangles.
(Here it is understood that a given triangulation, $T$, is counted
with the weight $1/|\mbox{Aut}(T)|$ where $|\mbox{Aut}(T)|$ is the order of its automorphism
group.) It is important to note that it is necessary to take into
account both of the terms $\Tr ({\cal ABC})$ and 
$\Tr ({\cal ACB})$ in~\rf{ZABC} if
one wishes to allow for any cyclic order of the colours around a given
triangle. If one considers only triangulations where the colours occur
in the same cyclic order for all triangles only one term is needed and
the model simplifies drastically. 
 The model with only one term, $\Tr({\cal ABC})$, was
studied by Cicuta et.\ al.\ who extracted numerically  the value of
the critical coupling constant and the critical index, 
$\gamma_{str}$~\cite{Cic93}. Later, it was shown that the model could
be mapped onto an $O(1)$ model on a random lattice with a
non-polynomial potential and this led to an exact solution of the 
model~\cite{CK96} showing that the critical indices of the model
coincide with those of the Ising model on a random
lattice. 

In stead of the model~\rf{ZABC} we shall consider the more general
model
\bea
{\cal Z}(g)&=&\int_{N\times N} \D A\prod_{i=1}^n \D B_i\,\D C_i \times
\nonumber \\
&& \exp\left\{ -N\Tr
\left[\frac{1}{g}V(A) +\frac{1}{2}\sum_{i=1}^n\left(B_i^2+C_i^2\right)
-\sum_{i=1}^n\left(
AB_iC_i+AC_iB_i\right)\right]\right\}.
\label{ZABCi}
\eea
This model is not more difficult to treat than the model~\rf{ZABC} and
reduces to the latter for $n=1$, and $V(A)=\frac{1}{2}A^2$, i.e.\footnote{
We note that in analogy with the $n=1$ case discussed above, for general $n$
the model~\rf{ZABCi} with only one term  $\sum_{i=1}^n\Tr(AB_iC_i)$ 
term can be mapped onto an  
$O(n)$
model with a non-polynomial potential and solved exactly. In particular, the
critical indices of the model can be shown to coincide with those of the
$O(n)$ model~\cite{CK96}.}
\beq
Z(g)=\left(\sqrt{g}\,\right)^{N^2}{\cal Z}(g)\left|_{n=1, 
V(A)=\frac{1}{2}A^2}\right.. 
\eeq
Integrating over the $C$-matrices in~\rf{ZABCi} we get
\beq
{\cal Z}(g)=\int_{N\times N} \D A\prod_{i=1}^n \D B_i\exp\left\{ -N\Tr
\left[\frac{1}{g}V(A) +\frac{1}{2}\sum_{i=1}^nB_i^2-\sum_{i=1}^n\left(
AB_iAB_i+A^2B_i^2\right)\right]\right\}.
\label{ZAB}
\eeq
In this letter we shall concentrate on solving the counting problem
for triangulations of spherical topology. We will derive an equation
which allows us to calculate, in an efficient way, the genus zero
contribution to all correlators of the type $\langle \frac{1}{N}\Tr
A^n\rangle$ iteratively in $g$. It will be clear, however, how to
extend the idea to obtain the higher genera contributions. We note
that just knowing $\langle \frac{1}{N}\Tr A^2\rangle$ is enough to
solve the three-colour problem, since we have for $n=1$, 
$V(A)=\frac{1}{2}A^2$
\beq
\langle \frac{1}{N}\Tr A^2\rangle =\frac{2g^2}{N^2}
\frac{d}{dg}\log {\cal Z}(g)=2g^2\frac{d}{dg}
\left\{\frac{1}{2}\log(g) +\frac{1}{N^2}\log Z(g)\right\}
\eeq
and in particular for genus zero (cf.\ equation~\rf{count})
\beq
\langle \frac{1}{N}\Tr A^2\rangle_{h=0}=
g+2g\sum_{v=0}^{\infty}N_0(2v)\,v \,g^v\equiv gT_1.
\label{countsolution}
\eeq
Baxter solved the regular lattice three-colour problem by a transfer
matrix method. Recently a transfer matrix formalism for random
triangulations has been invented~\cite{transfer} and it is natural to
ask whether this formalism can be applied to the present model. It
indeed can. A version of the transfer matrix formalism on random
lattices, applicable to loop gas models, was presented in~\cite{AKW97}
and can be applied to the three-colour problem exploiting its
equivalence with a variant of the $O(2)$ model on a random lattice. In
this approach no reference to any matrix model description is needed.
However, we shall stay within the matrix model approach because this
gives a faster way of deriving the equations we need.

In the subsequent section we shall write down the saddle point
equation corresponding to the matrix integral~\rf{ZAB}. This exposes
the analyticity structure of the model and shows why the full
three-colour problem is so much more complicated than the
restricted one. Actually one can derive the equation we are ultimately
after entirely by analyticity arguments, based on the saddle point
equation but we shall evoke another line of action, namely the loop
equation method. This method has the advantage that it is immediate to
see how to proceed to higher genera.

\newsection{The saddle point equation}
Integrating over the $B$- matrices in~\rf{ZAB} we get
\beq\label{ZA}
{\cal Z}(g)=\int_{N\times N}\, \D A\, \e^{-{N\over g}\Tr V(A) } \, \left( 
\mbox{det}(1\tens 1+ 1\tens A + A\tens 1 )
(1\tens 1 - 1\tens A - A\tens 1 )\right)^{-{n\over 2}}.
\eeq
Furthermore, integrating over the angular degrees of freedom leaves us
with the following integral over the eigenvalues of the matrix $A$
\beq\label{Zv.p.}
{\cal Z}(g)=\int\, \prod_{i=1}^N \D \l_i \: \prod_{j< k}
\left(\l_j-\l_k\right)^2\: \prod_{l,m} 
\left( (1+ \l_l + \l_m )( 1-\l_l-\l_m )\right)^{-{n\over 2}} \: 
\e^{-{N\over g} \sum_iV(\l_i) }.
\eeq
In the limit $N\rightarrow \infty$ the eigenvalue configuration is determined
by the saddle point of the integral above. The corresponding saddle point 
equation reads
\beq\label{sadpointij}
2 \sum_{j\neq i} {1\over \l_i-\l_j} \: - n \sum_{j} \left[ {1\over 1+\l_i+\l_j} - {1\over 1-\l_i-\l_j} \right] \: = {1\over g}V'(\l_i) \hspace{1.0cm}
\forall\, i=1\ldots N.
\eeq
Following reference~\cite{BIPZ78} we can introduce an eigenvalue density
$\rho(z)=\frac{1}{N}\sum_i\delta(z-\lambda_i)$ which in the limit 
$N\rightarrow\infty$ becomes a continuous function, and a resolvant 
$W(z)=\frac{1}{N}\sum_i\frac{1}{z-\lambda_i}\rightarrow \int d\lambda\,
\frac{\rho(\lambda)}{z-\lambda}$. As is clear from the
expression~\rf{Zv.p.}, the model becomes singular if one of the
eigenvalues approaches $-\frac{1}{2}$ or $\frac{1}{2}$. The solution
we are interested in corresponds to the situation where the eigenvalues
live on only one interval $[a,b]\subset [-\frac{1}{2},\frac{1}{2}]$ or
equivalently the situation where $W(z)$ is analytic in the complex
plane except for a cut $[a,b]\subset
[-\frac{1}{2},\frac{1}{2}]$. Written in terms of the resolvant the
saddle point equation reads
\beq\label{sadpoint}
W(z+i0)+W(z-i0)+n\left\{ W(1-z)+W(-1-z) \right\}\, = {1\over
g}V'(z); \hs 
z\in [a,b].
\eeq
This equation describes how $W(z)$ transforms when $z$ crosses the cut
and enters into another sheet. In the second sheet, $W(z)$ is a
combination of $W(z)$, $W(1-z)$ and $W(-1-z)$.
Thus  in the second sheet
there are three cuts: $[a,b]$, $[-1-b,-1-a]$ and $[1-b,1-a]$.
Crossing again these cuts, we generate an increasing number of cuts in
the next sheets, which means that $W(z)$ is defined on a Riemann
surface of infinite genus and with an infinite number of cuts in each
sheet. As opposed to this, for the 1-matrix model ($n=0$) $W(z)$ has
only one cut and two sheets and for the $O(n)$ model on a random
lattice  $W(z)$ has only two cuts in each sheet~\cite{Kos89}.
Here we need to consider and infinite series of cuts $\{I_k\}$ given
by
\beq
I_k=[a^{(k)},b^{(k)}], \hs\hs a^{(k)}=k+(-1)^k a,\hs b^{(k)}=k+(-1)^k b 
\eeq
and we note that the critical situation referred to above, where one of
the eigenvalues approaches $-\frac{1}{2}$ or $\frac{1}{2}$, corresponds
to the situation where all the cuts merge. One might expect, in
analogy with what was the case for the $O(n)$ model on a random
lattice, that this type of critical behaviour can only be realized for
a certain range of $n$-values~\cite{Kos89}. 
We note that when this type of critical
behaviour {\it is} realized the scaling behaviour of the eigenvalue
distribution in the vicinity of the endpoints of its support will be
as for the $O(n)$ model. For instance, if we consider $\rho(z)$ 
or equivalently $W(z)$ in the vicinity of $z=\frac{1}{2}$ we can view
the term $W(-1-z)$ as being regular and the saddle point equation
reduces to that of the $O(n)$ model on a random lattice.

Let us now show how it is possible, using the saddle point equation,
to build from $W(z)$ a function which has no cut at all. First we define 
\beq z_k=(-1)^k (z-k) \eeq
Then we can write the saddle point equation as
\beq
W(z_k+i0)+W(z_k-i0)+n\left\{ W(z_{k+1})+W(z_{k-1}) \right\}\, 
= {1\over g}V'(z_k), \hs  z\in [a_k,b_k]. 
\eeq
Multiplying this equation by $W(z_k+i0)-W(z_k-i0)$ we find that the function
\beq\label{sadptsq}
g(z)=
W^2(z_k)+nW(z_k)\left\{ W(z_{k+1})+W(z_{k-1}) \right\}\, 
- {1\over g}W(z_k)V'(z_k)
\eeq
has no cut along $I_k$ (while it of course has cuts along $I_{k-1}$
and $I_{k+1}$). Now it is easy to see that the following function,
$S(z)$,has no cut at all
\beq\label{defS}
S(z)=\sum_{k=-\infty}^{\infty} \: W^2(z_k) + n W(z_k)W(z_{k+1}) - {1\over g}\left( W(z_k)V'(z_k) - R(z_k) \right).
\eeq
Here $R(z)$ is the polynomial part of $W(z)V'(z)$ which we have
subtracted in order to ensure that the sum converges. 
Since $S(z)$ can have no singularities apart from the above
mentioned cuts it must be analytic in the whole complex
plane. Furthermore, it is easy to see that $S(z)$ fulfils the
following relations
\beq
S(1-z)=S(z), \hs\hs
S(z+2)=S(z).
\eeq
{}From the periodicity relation it follows that $S(z)$ can be written as
a Fourier series
\beq\label{Sp}
 S(z)= \sum_p S_p \,e^{i\pi p z}, \hs \hs
S_p=\int_{-1/2}^{3/2} S(z)\,^{-i\pi p z} \D z.
\eeq
Now it is actually possible by pure analyticity arguments to show
that $S_p=0$ $\forall p$, i.e.\ that $S(z)$ is identically equal to
zero. However, we shall derive this
result by another method, namely by means of the loop equations of the
model. These equation have the advantage that they contain also
information about higher genera contributions.

\newsection{The loop equations}
The loop equations simply express the invariance of the matrix
integral~\rf{ZAB} under analytic redefinitions of the integration 
variables. Let us introduce the notation $\vec{B}=(B_1,\ldots,B_n)$
and let us define
\begin{eqnarray}
W(z)  =  {1\over N}\LA \Tr {1\over z-A} \RA, &&
W(x,z)=\LA \Tr \frac{1}{x-A}\Tr \frac{1}{z-A}\RA_{conn},\\
W_2(z)  =  {1\over N}\LA \Tr {1\over z-A}\vec{B}^2 \RA, &&
T_B  =  {1\over N}\LA\Tr  \vec{B}^2  \RA, \\
H(z)  =  {1\over N}\LA \Tr {1\over z-A} \vec{B} A \vec{B} \RA, &&
F(z,x)  =  {1\over N}\LA\Tr {1\over z-A} \vec{B} {1\over x-A} \vec{B}
\RA
\end{eqnarray}
where the subscript $conn$ refers to the connected part. We note that
the leading contribution to all of the above listed correlators
is of the order $N^0$. Now, let us perform the following redefinition
of the field $A$
\beq
A\to A+\eps {1\over z-A}. \nonumber
\eeq
This gives rise to the identity
\beq
{1\over g}\left( V'(z)W(z)-R(z)\right) = 2 H(z)+2 zW_2(z) -2 T_B +
W^2(z)
+\frac{1}{N^2}W(z,z).
\label{loop1}
\eeq
Furthermore, considering the following redefinition of the field
$\vec{B}$ 
\beq
\vec{B}\to \vec{B}+\eps {1\over z-A}\,\vec{B}\,{1\over x-A} 
\nonumber
\eeq
we find
\begin{eqnarray}
\hspace{-0.5cm}\lefteqn{H(z)+H(x)+(x+2z)W_2(z)+(z+2x)W_2(x)-2 T_B}
\nonumber \\
& &\hspace{1.0cm}+ (1-(z+x)^2)F(z,x) - nW(z)W(x) -\frac{n}{N^2}W(z,x)= 0.
\label{loop2}
\end{eqnarray}
We see that as announced the loop equations (as usual) permit a study
of higher genera contributions. However, in the following discussion
we shall restrict ourselves to genus zero. First, we note that
equation~\rf{loop2} simplifies considerably  when $z+x=\pm 1 \equiv
\delta$.
Then, setting $x=\delta-z$ and combining~\rf{loop1} and~\rf{loop2} we
find 
\begin{eqnarray}
2\delta\left( W_2(z)+W_2(\delta-z)\right)&=&
-{1\over g}\left( V'(z)W(z)-R(z) +
V'(\delta-z)W(\delta-z)-R(\delta-z)\right)\nonumber\\
&&+
W^2(z)+2nW(z)W(\delta-z)+W^2(\delta-z) 
\end{eqnarray}
Replacing $z$ by $z_k$, and $\delta$ by $(-1)^k$ we have 
$\delta-z_k=z_{k+1}$ and summing over $k$ (from $k_1$ to $k_2-1$) we get
\bea
\lefteqn{\hspace{-2.0cm}
\sum_{k=k_1}^{k_2-1} \: \left[ W^2(z_k)+nW(z_k)W(z_{k+1})
-{1\over g}\left( V'(z_k)W(z_k)-R(z_k) \right) \right ] = }\nonumber
\\
&&
\hs \mbox{  }\oh \left( W^2(z_{k_1}) - {1\over g}\left( V'(z_{k_1})W(z_{k_1})-R(z_{k_1}) \right) \right) \nonumber \\
&&\hs - \oh \left( W^2(z_{k_2}) - {1\over g}\left( V'(z_{k_2})W(z_{k_2})-R(z_{k_2}) \right) \right) \nonumber \\
&&\hs +  \left( (-1)^{k_1} W_2(z_{k_1})- (-1)^{k_2} W_2(z_{k_2}) \right).
\eea
The right hand side vanishes when $k_1\to -\infty$ and $k_2\to +\infty$, while the left hand side is precisely the function $S(z)$ of equation~\rf{defS}.
Therefore we have
\beq S(z)=0. \label{Sz}\eeq
We thus have a closed (non-local) equation for $W$, which could in 
principle allow us to retrieve $W(z)$.

\newsection{An equation for the moments}
We shall now write equation~\rf{Sz} in a more manageable form, namely as an
equation for the moments $t_m=\LA \frac{1}{N}Tr A^m\RA$, related to $W(z)$ by
$$ W(z)\,{\mathop\sim_{z\to \infty}^{} } \, \sum_{m=0}^{\infty} \, 
{t_m\over z^{m+1}}. $$
We shall use the normalisation condition $t_0=1$ which corresponds to requiring
that the eigenvalue distribution is normalised to one.

First, we note that we can write
\beq
S(z)=\sum_{k=0}^{\infty}f(z_k) \label{Szk}
\eeq
with
\beq\label{fzk}
f(z)=W^2(z)+{n\over2}W(z)(W(1-z)+W(-1-z))-{1\over g}(V'(z)W(z)-R(z)).
\eeq
Next, we introduce the following functions
\beq 
\zeta_m(z)=\sum_k {1\over z_k^m},
\eeq
\beq
 \zeta_{m,p}(z)=\sum_k {1\over z_k^m}{1\over (1-z_k)^p},
\hspace{0.8cm} 
 {\overline\zeta}_{m,p}(z)=\sum_k {1\over z_k^m}{1\over (-1-z_k)^p} 
\eeq
The functions $\zeta_m(z)$ can be determined explicitly (by Fourier transform).
One has
\bea 
&& \zeta_1(z)={\pi\over \sin{\pi z}},\qquad \zeta_2(z)=
{\pi^2\over \sin^2{\pi z}}, \\
&& \zeta_{m+2}(z)= {1\over m(m+1)} {\D^2\over \D z^2}\zeta_m(z). \eea
In other words
\beq
 \zeta_m(z)=\left({\pi\over \sin{\pi z}}\right)^m \, P_m(\sin{\pi z})  
\eeq
where $P_m$  is a polynomial of degree less than or equal to $m-1$.
Actually, we can give an explicit expression for this polynomial, namely
$P_m(s)$ consists of the first $m-1$ powers of $s$  in
 the power series expansion
of $\left(\frac{s}{\arcsin s}\right)^m$  for $s$ in the vicinity of zero.
The functions $\zeta_{m,p}(z)$ and $\overline{\zeta}_{m,p}(z)$ can be expressed
in terms of the $\zeta_k(z)$ as follows 
\bea\label{zetarel1}
\zeta_{m,p}(z) & = & \sum_{k=1}^m \left(\begin{array}{c} m+p-1-k \\ p-1 
\end{array}\right) \zeta_k(z) + \sum_{k=1}^p \left(\begin{array}{c} m+p-1-k \\
 m-1 \end{array}\right) \zeta_k(z) \\
 {\overline\zeta}_{m,p}(z) & = & \sum_{k=1}^m \left(\begin{array}{c} m+p-1-k 
\\ p-1 \end{array}\right) (-1)^{m+p+k} \zeta_k(z) \nonumber\\
&&  \hspace{-0.5cm}\mbox{}+ \sum_{k=1}^p \left(\begin{array}{c} m+p-1-k \\ m-1 \end{array}\right) (-1)^{m+p+k} \zeta_k(z) \label{zetarel2}
\eea
Now, starting from the expression~\rf{Szk} and choosing the potential of our
model as
\beq
V(z)=\sum_j\frac{g_j}{j}z^j \label{pot}
\eeq
we can write
$$
S(z)=\sum_{k,l} t_k t_l \zeta_{k+l+2}\, +{n\over2}\sum_{k,l} t_k t_l 
\left( \zeta_{k+1,l+1} + {\overline\zeta}_{k+1,l+1}\right) \, -{1\over g}\left( \sum_{k=0}^{{\rm deg} V'}\sum_{l=k}^\infty g_k t_l \zeta_{l-k+1}  \right)=0.
$$
Using relations \rf{zetarel1} and~\rf{zetarel2}, and identifying the coefficients of each 
$\zeta_m$ we get for $m\geq 1$
\beq\label{eqtm}
{1\over g} \sum_{k=0}^{{\rm deg} V'} g_k t_{m+k-1}
=
\sum_{k=0}^{m-2} t_k t_{m-k-2} \, +2n \sum_{k=0}^\infty 
\sum_{l=0}^{2k+1} \left(\begin{array}{c} 2k+1 \\ 2k+1-l \end{array}\right) \,
 t_{m+l-1} t_{2k+1-l}.  
\eeq
This equation is equivalent to equation~\rf{Sz} and contains all information
about the full non-perturbative solution of the model. Lacking a means of
solving this equation exactly we shall in the next section describe how 
a perturbative solution can be found.

\newsection{Perturbative solution}
Let us specialise to the case of a Gaussian potential, i.e.\
$V'(z)=z$, but keep $n$ arbitrary. Setting $n=1$ at any stage of the
calculation then brings us to the three-colour problem. For a
quadratic potential, obviously $t_{2m+1}=0$, $\forall m$, and the
function $W(z)$ is an odd function of $z$ which is analytic in the
complex plane except for a cut of the type $[-a,a]\subset
[-\frac{1}{2},\frac{1}{2}]$. The relation~\rf{eqtm} in this case
reduces to 
\beq
{1\over g}t_{2m} = \sum_{k=0}^{m-1} t_{2k}t_{2(m-k-1)} \, + \, 
2n \sum_{k,l=0}^{\infty}\,
\left(\begin{array}{c} 2k+2l+1 \\ 2k+1\end{array}\right)
\, t_{2(m+k)}t_{2l}, \qquad \,\, m\geq 1.
\label{recursion0}
\eeq
Setting $t_{2m}=g^m T_m$ we have
\beq
T_m = \sum_{k=0}^{m-1} T_k T_{m-k-1} \, + \, 2n g
\sum_{k,l=0}^{\infty} 
\left(\begin{array}{c} 2k+2l+1 \\ 2k+1\end{array}\right)
g^{k+l} \, T_{m+k}T_l, 
\qquad \,\, m\geq 1.
\label{recursion1}
\eeq
This equation can be solved perturbatively in $g$: First we note that
for $g=0$, the model under consideration is nothing but the Gaussian
one-matrix model for which the function $W(z)$ is known to be
\beq
W^{(0)}(z)=\frac{1}{2g}\left( z-\sqrt{z^2-4g}\,\right).
\label{W0z}
\eeq
To the zeroth order in $g$ we hence have
\beq
 T^{(0)}_m=\tau_m={(2m) !\over m!\, (m+1)!}. 
\label{Tm0}
\eeq
Using as initial condition~\rf{Tm0} and the normalisation condition
$T_0=1$, it is obvious that equation~\rf{recursion1} 
allows one to calculate the
individual moments order by order in $g$. 
We remind the reader that just knowing the moment $T_1$ (for $n=1$) one
has the solution of the three-colour problem on a random lattice (cf.\
equation~\rf{countsolution}). Writing $T_1=\sum_{i=1}^\infty T_1^{(i)} g^i$ it
appears that to determine $T_1^{(i)}$ one needs to  calculate 
$\{T_p^{(q)}\}_{p=1,\ldots,i-1;q=1,\ldots,i-p}$. Below we give $T_1$ to
the first six orders in $g$
\bea
T_1^{(1)}&=&2n, \nonumber \\
T_1^{(2)}&=&10n+4n^2, \nonumber \\
T_1^{(3)}&=&70n+60n^2 +8n^3, \nonumber \\
T_1^{(4)}&=&588n+764n^2+240n^3 +16n^4, \nonumber \\
T_1^{(5)}&=&5544n+9520n^2+4840n^3+800n^4+32n^5,\nonumber \\
T_1^{(6)}&=&56628n+119704n^2+84216n^3+23440n^4+2400n^5 + 64n^6. \nonumber
\eea
Unfortunately, we have not been able to express $T_1(g)$ in a closed
form, neither for $n$ general, nor for $n=1$. 
However, we can do better than determining the moments individually. 
As we shall see, by an appropriate
ansatz we can replace the quadratic  recursion relation~\rf{recursion1} by
a linear one and solve simultaneously for all $T_i$. In stead of
solving the model iteratively in $g$ we shall solve it iteratively in
$a^2$ where $a$ is the endpoint of the cut of $W(z)$. It is obvious
from the eigenvalue integral~\rf{Zv.p.} as well as from the 
relation~\rf{W0z} that small values of $g$ correspond to small values
of $a^2$.

 Let us now set $t_{2m}={\cal T}_m d^{\,m}$ where  ${\cal T}_0=1$ and
$d=\left(\frac{a}{2}\right)^2$. Then
we get from~\rf{recursion0}
\beq
X{\cal T}_{m} = \sum_{l=0}^{m-1} {\cal T}_{l}\,{\cal T}_{m-l-1} \, + \, 
2n\, d\sum_{k,l=0}^{\infty}\,
\left(\begin{array}{c} 2k+2l+1 \\ 2k+1\end{array}\right)
\, {\cal T}_{m+k}{\cal T}_{l}\, d^{\,k+l},\qquad \,\, m\geq 1,
\label{drecursion}
\eeq
where
\beq
X=\frac{d}{g}=\frac{a^2}{4g}.\label{endpoint}
\eeq
Equation~\rf{drecursion} must be supplemented by the following boundary
condition
\beq
\frac{{\cal T}_m}{\tau_m}\sim const. 
\hspace{0.5cm}\mbox{as}\hspace{0.5cm} m\rightarrow \infty.
\label{boundary}
\eeq
 This relation expresses the fact that $W(z)$ has a
square root branch point at $z=a$.
Equation~\rf{drecursion} can be solved iteratively in $d$. 
To leading order in $d$ the solution is again given by~\rf{W0z},
i.e.\ we have
\beq
{\cal T}_m^{(0)}=\tau_m,\hspace{1.0cm}X^{(0)}=1.
\eeq
Based on an analysis of the structure of the solution found after the
first steps of the iteration process we introduce the following ansatz
\bea
{\cal T}_m&=&\sum_{j=0}^\infty \tau_{m+j}\,v_j\, d^{\,j}\,,
\label{ansatz1} \\
X&=&\sum_{j=0}^{\infty}X^{(j)}\, d^{\,j},
\hspace{1.0cm} v_j=\sum_{i=0}^{\infty}v_j^{(i)}\, d^{\,i}
\label{ansatz2}
\eea
with
\beq
v_0^{(0)}=1,\hspace{1.0cm}  v_s^{(0)}=0,\:\: s\neq 0.
\eeq
This ansatz fulfills the boundary condition~\rf{boundary} to any finite 
order in $d$ and the requirement that
equation~\rf{drecursion} should be satisfied and that ${\cal T}_0=1$
will give us a linear 
recursion relation for the $v_j$'s. We note that
the  function $X=X(d)$ encodes the information about the
critical behaviour of the model. Critical behaviour occurs when the
radius of convergence of the power series $X=X(d)$ is reached. We
know that we will encounter one type of singularity when $d$
approaches $\frac{1}{16}$ but there might be others.
Inserting the ansatz~\rf{ansatz1} in~\rf{drecursion} we get
\bea
\sum_{s=0}^{\infty}\tau_{m+s}\,d^{\,s}\left\{Xv_s-\sum_{q=0}^sv_q v_{s-q}+
2\sum_{q=0}^s \sum_{l=0}^\infty
v_{q+l+1}v_{s-q}\,\tau_l\, d^{\,l+1}\hspace{1.0cm}\right.
&& \nonumber \\
\left. -2nd \sum_{q=0}^s\sum_{l,b=0}^{\infty}
v_{s-q}v_b\, \tau_{l+b}\,d^{\,l+b} 
\left(\begin{array}{c} 2q+2l+1 \\ 2q+1\end{array}\right)\right\}&=&0.
\eea
Since this equation must be fulfilled for any value of $m$ and to any order
in $d$
 the quantity in the curly bracket must vanish. This
implies that for all $s$ we have
\bea
\sum_{q=0}^s v_{s-q}\left\{\delta_{q,0}\,\,X -v_q +
2\sum_{l=0}^\infty v_{q+l+1}\,\tau_l\, d^{\,l+1} 
\hspace{3.0cm}\right.\nonumber &&\\
\left. \mbox{}-2nd \sum_{b,l=0}^{\infty} d^{\,l+b}v_b\,\tau_{l+b}\,
\left( \begin{array}{c} 2q+2l+1 \\ 2q+1\end{array}\right)\right\}&=&0.
\eea
Since $v_0\neq 0$ this equation can only be satisfied for all $s$ if
the term in
the curly bracket vanishes for all $q$, i.e.
\beq
\delta_{q,0} X -v_q +2\sum_{l=0}^{\infty} v_{q+l+1}\,\tau_{l}\,d^{\,l+1}
-2nd\sum_{b,l=0}^{\infty} v_b\, \tau_{l+b}\,d^{\,l+b} 
\left(\begin{array}{c} 2q+2l+1 \\ 2q+1 \end{array}\right)=0.
\label{recursion}
\eeq
Now we have succeeded in replacing the quadratic recursion
relation~\rf{eqtm} by a linear one! The relation~\rf{recursion} must
be supplemented by the normalisation condition
\beq
T_0=\sum_{i=0}^{\infty}\,\tau_i\, v_i\, d^{\,i}=1. 
\label{norm}
\eeq
From~\rf{recursion} and~\rf{norm} we can determine $X$,
$\{v_i\}_{i=1}^{\infty}$ and hence $W(z)$ to (in principle) any order
in $d$. 
Inserting the expansion~\rf{ansatz2} in~\rf{recursion}
and~\rf{norm}
we get
\beq
v_0^{(i)}=-\sum_{j=1}^{i-1}v_j^{(i-j)}\,\tau_j,
\eeq
\bea
X  ^{(i)}\,\delta_{q,0}-v_q^{(i)}+2
\sum_{j=1}^{i-1}v_{q+i-j}^{(j)}\,\tau_{i-j-1}\hspace{6.0cm}&& \nonumber \\
-2n\sum_{l=0}^{i-1}\,\sum_{p=0}^{i-l-1}v_{i-l-p-1}^{(p)}\,\tau_{i-p-1}
\left(\begin{array}{c} 2q+2l+1 \\ 2q+1 \end{array} \right)=0
\eea
where on our way we have made use of the initial condition
$v_i^{(0)}=0$ for $i\neq 0$.
 To solve recursively these
equations we determine in each step first $v_0^{(i)}$, next $X^{(i)}$
and then $v_j^{(i)}$, for $j>0$. We note that in order to determine $X(d)$
and $W(z)$ to order $s$ in $d$ we need only to know
$\{v_i^{(j)}\}_{i=0,\ldots s;j=0,\ldots, s-i}$. 
Since by our iteration process we find indeed a perturbative solution
of equation~\rf{drecursion} and since this equation is known to have a unique 
perturbative solution our ansatz is justified.
Below we give the
function $X(d)$ to the first seven orders in $d$
\bea
X^{(1)}&=&2n, \nonumber\\
X^{(2)}&=&12n, \nonumber \\
X^{(3)}&=&100n, \nonumber \\
X^{(4)}&=&980n-12n^2, \nonumber \\
X^{(5)}&=&10584n-360n^2, \nonumber \\
X^{(6)}&=&121968n-7424n^2,\nonumber \\
X^{(7)}&=&1472328n-132160n^2+240n^3. \nonumber 
\eea
We remind the reader that the function $X=X(d)$ is related to the
endpoint of the cut of $W(z)$ by the relation~\rf{endpoint} and that
the three-colour problem corresponds to the case $n=1$. Unfortunately,
we have not been able to express $X(d)$ in a closed form, neither for 
$n$ general, nor for $n=1$. However, we recognise the coefficient of the
term linear in $n$ as the integral of an elliptic function~\cite{Han75}
\beq
2\,d+12\,d^2+100\,d^3+980\,d^4+\ldots=
\sum_{k=1}^{\infty}\frac{\left[(2k)!\right]^2}{(k!)^3(k+1)!}\, d^k
=\frac{1}{8\pi d}\int_{0}^{16d}K(t^{1/2})\,dt\,-\,1 \label{int}
\eeq
We here explicitly see that the model becomes singular when $d$ approaches
$\frac{1}{16}$ as argued earlier. By means of our results for $X(d)$ we
can determine the corresponding value of the coupling constant, namely
\beq
\frac{1}{g_c}=\frac{X(d_c)}{d_c}, \hspace{0.7cm}d_c=\frac{1}{16}.
\eeq
In particular, using~\rf{int}, we find to the first order in $n$
\beq
\frac{1}{g_c}=16\left\{1+n\left(\frac{4}{\pi}-1\right)+
{\cal O}(n^2)\right\}.
\eeq

\newsection{Conclusion}

 Exploiting the matrix model formulation of the three-colour problem on 
a random lattice we have developed an algorithm which allows us to solve the
problem recursively. In addition we have exposed the analyticity structure
of the problem and argued why the full three-colour problem is so much more
difficult to solve than its restricted counterpart. 

Let us finish by mentioning an interesting observation. By arguments
analogous to those given in the introduction for the three-colour
model it can be seen that the model given by~\rf{ZABCi} is equivalent
to an $O(2n)$ model on a random lattice where the loops are restricted
to having even length. We have seen that the critical behaviour of the
model~\rf{ZABCi} is the same as that of the $O(n)$ model on a random
lattice. Hence, restricting the loops to be of even length takes the
$O(n)$ model to the universality class of the $O(\frac{n}{2})$ model.

\vspace{15pt}
\noindent
{\bf Acknowledgements} \hspace{0.3cm} 
It is a pleasure to thank E.\ Guitter and M.\ Harris for
interesting discussions and L.\ Chekhov for participating in the first 
stages of this project. In addition B.\ Eynard acknowledges the support
of the European Union through their TMR programme, 
contract No.\ ERBFMRXCT 960012.

\end{document}